\documentclass[a4paper,aps,prd,10pt,preprintnumbers,showpacs,twocolumn,superscriptaddress,nofootinbib,amsmath,amssymb]{revtex4-1}
\usepackage{cases}
\usepackage{graphicx}
\usepackage[utf8]{inputenc}
\usepackage[T1]{fontenc}
\usepackage{cmap}

\def\imo{i}

\begin{document}

\title{Stable Schwarzschild stars as black-hole mimickers}

\author{R. A. Konoplya}\email{roman.konoplya@gmail.com}
\affiliation{Institute of Physics and Research Centre of Theoretical Physics and Astrophysics, Faculty of Philosophy and Science, Silesian University in Opava, Bezručovo nám. 13, CZ-746 01 Opava, Czech Republic}
\affiliation{Peoples Friendship University of Russia (RUDN University), 6 Miklukho-Maklaya Street, Moscow 117198, Russian Federation}
\author{C. Posada}
\email{camilo.posada@fpf.slu.cz}
\affiliation{Institute of Physics and Research Centre of Theoretical Physics and Astrophysics, Faculty of Philosophy and Science, Silesian University in Opava, Bezručovo nám. 13, CZ-746 01 Opava, Czech Republic}
\author{Z. Stuchlík}
\affiliation{Institute of Physics and Research Centre of Theoretical Physics and Astrophysics, Faculty of Philosophy and Science, Silesian University in Opava, Bezručovo nám. 13, CZ-746 01 Opava, Czech Republic}
\author{A. Zhidenko}\email{olexandr.zhydenko@ufabc.edu.br}
\affiliation{Institute of Physics and Research Centre of Theoretical Physics and Astrophysics, Faculty of Philosophy and Science, Silesian University in Opava, Bezručovo nám. 13, CZ-746 01 Opava, Czech Republic}
\affiliation{Centro de Matemática, Computação e Cognição (CMCC), Universidade Federal do ABC (UFABC),\\ Rua Abolição, CEP: 09210-180, Santo André, SP, Brazil}

\begin{abstract}
The Schwarzschild star is an ultracompact object beyond the Buchdahl limit, which has Schwarzschild geometry outside its surface and positive pressure in the external layer which vanishes at the surface. Recently it has been shown that the Schwarzschild star is stable against spherically symmetric perturbations. Here we study arbitrary axial nonspherical perturbations, and show that the observable quasinormal modes can be as close to the Schwarzschild limit as one wishes, what makes the Schwarzschild star a very good mimicker of a black hole. The decaying time-domain profiles prove that the Schwarzschild star is stable against nonspherical perturbations as well. Another peculiar feature is the absence of echoes at the end of the ringdown. Instead we observe a nonoscillating mode which might belong to the class of algebraically special modes. At asymptotically late times, Schwarzschildian power-law tails dominate in the signal.
\end{abstract}
\maketitle


\section{Introduction}\label{intro}

The detection of gravitational waves (GW) by the LIGO and Virgo Collaborations~\cite{abbott2016a,abbott2016} as well as recent observations of black holes in the electromagnetic spectrum \cite{Akiyama:2019cqa,Goddi:2017pfy} gave birth to a new era in the astronomy of astrophysical compact objects. It is commonly assumed that black holes and neutron stars are the most popular candidates for current and future detections of GW. However some authors found that the observations could also be interpreted as due to non-Einsteinian black holes \cite{Yunes:2016jcc,Konoplya:2016pmh} or even alternatives to black holes \cite{abbott2016,cardoso2016,cardoso2016b,abedi2017,Konoplya:2016hmd} or the so-called exotic compact objects (ECOs) (see \cite{cardoso2019} for a recent review), which have been proposed as solutions to clear up the several paradoxes of classical black holes in the Einstein theory \cite{wald2001}.

One of these ECOs is the gravastar proposed by Mazur and Mottola \cite{mazur2001,mazur2004}, which has attracted considerable interest in the last decade (see, for instance, \cite{cattoen2005,benedictis2005,lobo2006,cardoso2008,chirenti2008,sakai2014,pani2015,volkel2017,Nakao:2018knn,Wang:2018cum} and references therein).  In its original form a gravastar is a four-layer configuration which consists of an interior region with equation of state (EOS) $p=-\epsilon<0$, described by a patch of the de Sitter metric, which is matched through a boundary layer to a shell filled with a stiff ultrarelativistic fluid with EOS $p=\epsilon$. Likewise, there is a second boundary layer that matches the shell to the exterior Schwarzschild geometry. The gravastar is a nonsingular solution without an event horizon, therefore providing a legitimate alternative to black holes.

The issue of stability of gravastars has been thoroughly investigated. Visser and Wiltshire~\cite{visser2004} studied the radial stability of a simple three-layer gravastar model, and Horvat {\it et al.} \cite{horvat2011} considered the stability of a continuous pressure gravastar. In a seminal paper, Chirenti and Rezzolla~\cite{chirenti2007} investigated the stability of gravastars against axial perturbations  and computed their quasi-normal modes (QNMs). They concluded that  the spectrum of axial QNMs of gravastars is different from those of a Schwarzschild black hole. Similar conclusions were drawn by Pani {\it et al} \cite{pani2009} who computed axial and polar oscillations of thin-shell gravastars. Scalar perturbations of gravastars were studied \cite{cardoso2008,chirenti2008} in the context of the ergoregion instability. More recently, Cardoso {\it et al} \cite{cardoso2014} studied linear perturbations in the gravastar spacetime. In \cite{chirenti2016}, it was argued that the gravastars are discarded by the GW150914 observations, though there the mass of the gravastar was fixed by that value which follows from the supposition that the central object under consideration is the Kerr black hole. After that, a kind of extrapolation of the non-rotating case was done via adding terms from quasinormal modes of the slowly rotating neutron star \cite{chirenti2016}. This way, a large existing uncertainties (of tens of percents) in the resultant angular momentum and mass were not taken into consideration in \cite{chirenti2016}.

More recently, in connection with gravastars, Mazur and Mottola \cite{mazur2015} reconsidered the Schwarzschild interior solution \cite{schwarzschild1916b} or Schwarzschild star, which is an exact solution of the Einstein's equations for a spherically symmetric mass with uniform energy density. It is well known that when the radius of the Schwarzschild star reaches the Buchdahl bound $R=(9/4)M$, the central pressure diverges. The existence of this limiting configuration is not restricted to the case of constant density stars. Buchdahl \cite{buchdahl1959} showed that if one assumes a positive energy density which decreases monotonically with $r$, then there is a general upper mass bound $M\leq 4R/9$ which is independent of the relation between pressure and density.
Configurations more compact than those within the Buchdahl limit were believed to be unstable, as one would have either assume the existence of the singularity inside the star or the pressure which is increasing outward. The uniform density configurations were studied in \cite{Boehmer:2003uz,Stuchlik:2008xe,Stuchlik:2016xiq,Stuchlik:2017qiz}.
A model of black-hole mimicker in the form of a nonlocal star in the context of ghost-free, infinite derivative gravity was proposed in \cite{Buoninfante:2019swn}.

Therefore, the existence of the Buchdahl limit together with the assumption of constant density, considered `unphysical' (see however \cite{misner1973}), has left the Schwarzschild star unstudied thoroughly (a notable exception can be found in \cite{cattoen2005}). Nevertheless, when one does consider the constant density star below the Buchdahl limit, $2M<R<(9/4)M$, a compelling behavior of the solution can be observed \cite{cattoen2005,mazur2015}. For instance, the pole where the pressure is divergent moves outward, starting at the center of the star, up to a surface of radius $R_{0}=3R\sqrt{1-\frac{8}{9}\frac{R}{R_{\rm S}}}<R$. Then a new region naturally emerges in the range $0<r<R_{0}$,  characterized by a negative pressure. In the limit when $R_{0}\to R_{\rm S}^{-}$ from the interior and $R\to R_{\rm S}^{+}$ from the exterior, the full interior region becomes one with constant negative pressure with the geometry of a modified de Sitter spacetime. The interior region is matched to the vacuum exterior Schwarzschild metric through a surface layer of transverse stresses endowed with a surface tension. The pressure in this exterior layer is positive. Thus, in the ultracompact limit
$R\to R_{S}$ the Schwarzschild star resembles the main features of the gravastar proposed in \cite{mazur2001,mazur2004}, although one with no thin-shell of matter.

The ultracompact Schwarzschild star was extended to slow rotation in \cite{posada2017}. The main conclusion of that work is that the moment of inertia and quadrupole moment, or I-Q relations, for a slowly rotating ultracompact Schwarzschild star approaches to the corresponding values for the Kerr metric. Recently Posada and Chirenti~\cite{posada2019} studied the radial stability of this configuration using the `pulsation' equation derived by Chandrasekhar \cite{chandrasekhar1964}. They found that the ultracompact Schwarzschild star is stable against radial perturbations. This result indicates, in principle, that the Schwarzschild star might be a viable model for mimicking the Schwarzschild black-hole behavior in the limit $R\to R_{S}$, when the geometry must resemble Schwarzschildian one up to the very thin layer near the surface of the star. Apparently, one should expect that for such spacetimes, the nearly Schwarzschild quasinormal ringing must end up with a series of echoes, which come from the modification near the surface \cite{cardoso2016,cardoso2016b,Konoplya:2018yrp}.

In the present paper we have found further evidences of the viability of the Schwarzschild star. First of all, we have shown the stability against nonspherical (axial) perturbations. The remarkable property of the axial spectrum of the Schwarzschild star is that the expected nearly Schwarzschild quasinormal modes end up not with echoes, but with a new nonoscillating decaying mode, which, in its turn, goes over into the asymptotic power-law tails. The latter are identical to the Schwarzschild ones. Thus, the $l>1$ perturbations of the Schwarzschild star are indistinguishable from those of the black hole neither at the time of quasinormal ringing, nor at later times, including the asymptotic tails.

This paper is organized as follows. In section~\ref{sec1} we review the Schwarzschild star and the emergence of a negative pressure interior when we consider ultracompact configurations beyond the Buchdahl limit. In section~\ref{sec2} we find quasinormal of axial gravitational perturbations via time-domain integration. Finally, in the conclusion we summarize the obtained results and mention some important open questions.


\section{Ultracompact Schwarzschild stars and the gravastar limit}\label{sec1}

In this section we briefly review the Schwarzschild interior solution~\cite{schwarzschild1916b}, or Schwarzschild star, corresponding to a spherical mass of constant energy density and the emergence of the interior region with negative pressure beyond the Buchdahl bound~\cite{mazur2015}. We begin with a spherically symmetric spacetime in the standard Schwarzschild coordinates \footnote{Throughout the paper we use geometrized units $G=c=1$. We shall measure all quantities in units of the Schwarzschild radius $R_{S}=2M$.}

\begin{equation}\label{metric}
ds^2 = -f(r)dt^2+\frac{dr^2}{h(r)}+r^2(d\theta^2+\sin^2\theta~d\phi^2),
\end{equation}
where the metric functions are given by \cite{misner1973}
\begin{numcases}{f(r) = }\label{interiorf}
\frac{1}{4}\left(3\sqrt{1-\frac{R_{S}}{R}}-\sqrt{1-\frac{R_{S}r^2}{R^3}}\right)^2\!\!\!,\!\!\!\!\!\! & $r < R$, \\
\left(1-\frac{R_{S}}{r}\right), & $r \geq R$;
\end{numcases}
and
\begin{numcases} {h(r) =}
\left(1-\frac{R_{S}r^2}{R^3}\right), & $r < R$, \label{interiorh} \\
\left(1-\frac{R_{S}}{r}\right), & $r \geq R$,
\end{numcases}
where $R_{S}=2M$ is the Schwarzschild radius, $M$ denotes the gravitational mass and $R$ the radius of the configuration. The relation for the pressure is

\begin{equation}\label{interiorp}
p(r) = \epsilon\cdot\frac{\sqrt{1-\dfrac{R_{S}r^2}{R^3}} - \sqrt{1-\dfrac{R_{S}}{R}}}{3\sqrt{1-\dfrac{R_{S}}{R}}-\sqrt{1-\dfrac{R_{S}r^2}{R^3}}},
\end{equation}
where $\epsilon$ denotes the mass-energy density, which is a constant,
\begin{equation}
\epsilon =\left\{
  \begin{array}{ll}
    \dfrac{3M}{4\pi R^3}=\dfrac{3R_{S}}{8\pi R^3}, & r < R, \\
    0, & r \geq R.
  \end{array}
\right.
\end{equation}

Notice that the pressure vanishes $p(R)=0$ at the surface $r=R$. The Schwarzschild star is matched at the boundary $r=R$, with the asymptotically flat exterior Schwarzschild solution. Note that the central pressure becomes infinite when the radius of the star reaches the Buchdahl bound $R=(9/8)R_{S}$~\cite{buchdahl1959}. However, a further analysis shows that \eqref{interiorp} is regular except at some radius $R_{0}$ where the denominator in \eqref{interiorp},
\begin{equation}\label{denom}
D \equiv 3\sqrt{1-\frac{R_{S}}{R}}-\sqrt{1-\frac{R_{S}r^2}{R^3}},
\end{equation}
vanishes in the range $0<r<R$.  From \eqref{interiorf} and \eqref{interiorp}, it can be observed that the pressure is divergent at the same point where $f(r)=0$. This singular radius can be found directly from \eqref{denom} to be
\begin{equation}\label{r0}
R_{0} = 3R\sqrt{1-\frac{8}{9}\frac{R}{R_{S}}},
\end{equation}
which is complex for $R>(9/8)R_{S}$. When the radius of the star reaches the Buchdahl bound, $R_{0}=0$ which indicates that the pressure diverges \emph{first} at the center. If one considers the Schwarzschild star in the regime, $R_{S}<R\leq9R_{S}/8$, it can be observed that the divergence in the pressure \emph{moves out from the center} to a surface of radius $R_{0}$ given by \eqref{r0}, which is defined in the range $0<R_{0}<R$. Note that in the regime beyond the Buchdahl bound, the denominator \eqref{denom} changes sign; therefore, a new \emph{nonsingular} solution with \emph{negative} pressure emerges naturally in the region $0\leq r < R_{0}$. On the other hand, the interior metric component $-g_{tt}=f(r)$ given by \eqref{interiorf} is a perfect square, thus, it remains positive.

From~\eqref{interiorf}~and~\eqref{interiorh}, we note that in the `black-hole' limit when $R=R_{0}=R_{S}$, the complete interior region becomes a patch of modified de Sitter metric

\begin{equation}\label{metric0}
f(r) = \frac{1}{4}h(r),\quad h(r)=1-\left(\frac{r}{R_{S}}\right)^2,\quad r < R=R_{S}
\end{equation}
with constant negative pressure $p=-\epsilon$ corresponding to the `dark energy' EOS. It is relevant to remark that the factor $1/4$ in the $g_{tt}$ metric component modifies the standard de Sitter metric, and it is essential for the proper matching of the Schwarzschild star with the exterior Schwarzschild metric.

Note that in contrast to the model proposed originally in~\cite{mazur2001,mazur2004}, here the phase transition occurs first at the center, once the Buchdahl bound is reached. The surface $R=R_{S}$ is a null surface; however, there is no interior trapped surface and \emph{there is no} formation of an event horizon. The divergence in the pressure at $R_{0}$ can be integrated through the Komar formula~\cite{poisson2004}, if one assumes the presence of a transverse pressure at the hypersurface $R_{0}$, such that

\begin{equation}\label{delta}
8\pi\sqrt{\frac{f}{h}}r^2(p_{\perp}-p)=\frac{8\pi\epsilon}{3} R_{0}^3\delta(r-R_{0}).
\end{equation}
This $\delta$ function, associated to a surface layer of transverse stresses, provides a strict matching at the surface as required by the Israel junction conditions~\cite{poisson2004,israel1966} on null surfaces~\cite{barrabes1991} (see~\cite{mazur2015} for details).

Note that in contrast with the model proposed in~\cite{mazur2001,mazur2004} where the authors introduced a thin shell filled with a stiff fluid $p=\epsilon$, to join the interior de Sitter with the Schwarzschild exterior, in this new description there is no thin shell of matter. Instead, the strict matching between the ultracompact Schwarzschild star and the exterior Schwarzschild metric can be carried out if one relaxes the isotropic pressure condition $p=p_{\perp}$, originally assumed in~\cite{schwarzschild1916b}, and allows the presence of a boundary layer of transverse stresses $p\neq p_{\perp}$ at $r=R_{0}$, with an effective zero thickness, endowed with a surface tension given by~\cite{mazur2015}

\begin{equation}\label{tsurf}
\tau_{s}=\frac{MR_{0}}{4\pi R^3}=\frac{\Delta\kappa}{8\pi},
\end{equation}
which is proportional to the difference in magnitude of the surface gravities
\begin{equation}\label{kappa}
\Delta\kappa=\kappa^{+}-\kappa^{-}=\frac{R_{S}R_{0}}{R^3}.
\end{equation}

This is the assumption which provides the physical viability of the Schwarzschild star beyond the Buchdahl limit and the \emph{natural emergence} of a negative pressure interior. It is worthwhile to remark that all these quantities described above are related to purely mechanical effects in sharp contrast with the classical analog of black holes mechanics and thermodynamics~\cite{bekenstein1973,bardeen1973}. For instance, surface area is related solely to the surface area of the configuration and not to entropy; surface gravity is surface tension and not temperature. Furthermore, this configuration has zero entropy and temperature, indicating a condensate state. Finally, the Schwarzschild star,  near the Schwarzschild regime $R\to R_{S}$, provides a nonsingular external surface with no event horizon, which makes it viable as a black-hole 'mimicker'. In the next section we will study its stability against axial gravitational perturbations.

\section{Gravitational perturbations of ultracompact Schwarzschild stars}\label{sec2}

\begin{figure*}
\includegraphics[width=.5\linewidth]{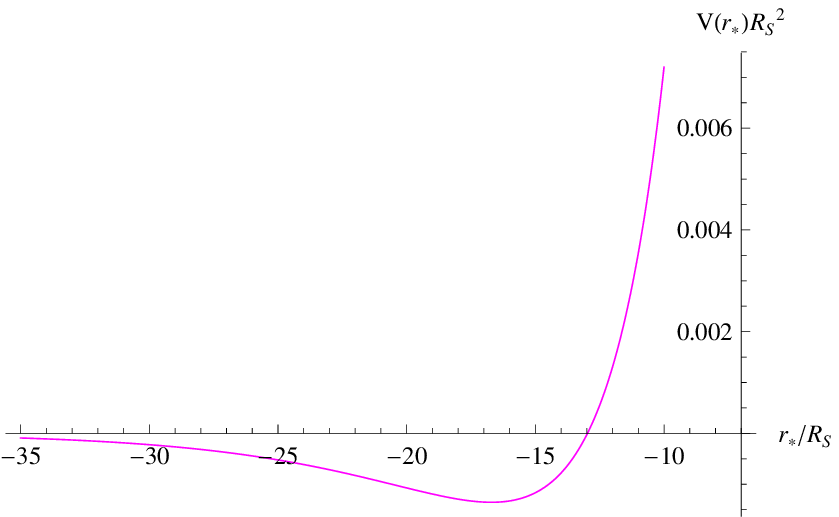}\includegraphics[width=.5\linewidth]{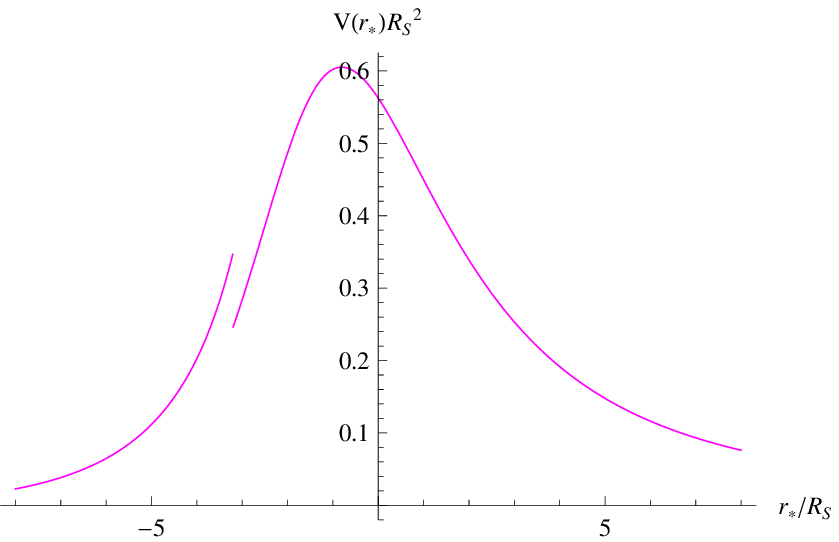}
\caption{Effective potential for the axial gravitational perturbations ($l=2$) of Schwarzschild star $R=1.1R_{S}$. Left panel: negative gap when approaching the singular point $r\to R_{0}$. Right panel: discontinuity at $r=R$.}\label{fig:potp}
\end{figure*}

\begin{figure}
\includegraphics[width=\linewidth]{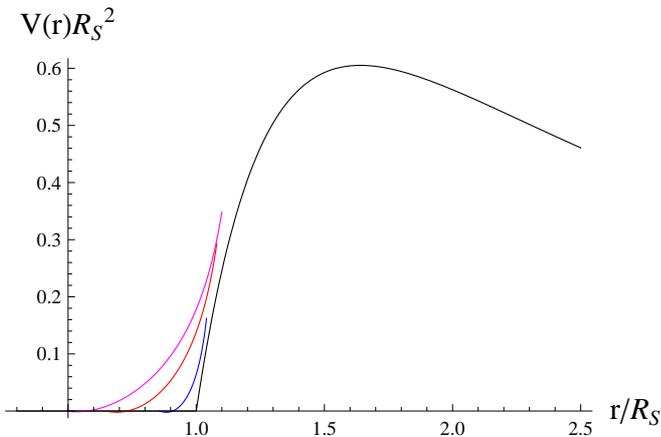}
\caption{Effective potentials for the axial gravitational perturbations ($l=2$) of the Schwarzschild black hole (black) and for the interior ($R_{0}<r<R$) of Schwarzschild stars: $R=1.04R_{S}$ (blue, lower), $R=1.08R_{S}$ (red), $R=1.1R_{S}$ (magenta, upper).}\label{fig:potentials}
\end{figure}

The axial gravitational perturbations for a relativistic mass, with given mass-energy density $\epsilon$ and pressure $p$, are determined by the master equation

\begin{equation}\label{master}
\left(\frac{\partial^2}{\partial t^2}-\frac{\partial^2}{\partial r_{*}^2}+V_{l}(r)\right)\Psi(r,t)=0
\end{equation}
where the effective potential $V_{l}$ is given by~\cite{chandra1991a,chandra1991b}
\begin{equation}\label{potential}
V_{l}(r) = f(r)\left(\frac{l(l+1)}{r^2}-6\frac{m(r)}{r^3}-4\pi p(r)+4\pi\epsilon\right),
\end{equation}
where $m(r)$ is the mass enclosed by $r$
\begin{equation}
m(r) = 4\pi\int_{0}^{r}\epsilon\cdot r^2dr,
\end{equation}
and $r_{*}$ denotes the ‘tortoise' coordinate defined in the standard form by
\begin{equation}\label{tortoise}
dr_{*} = \frac{dr}{\sqrt{f(r)h(r)}}.
\end{equation}

\begin{figure}
\includegraphics[width=\linewidth]{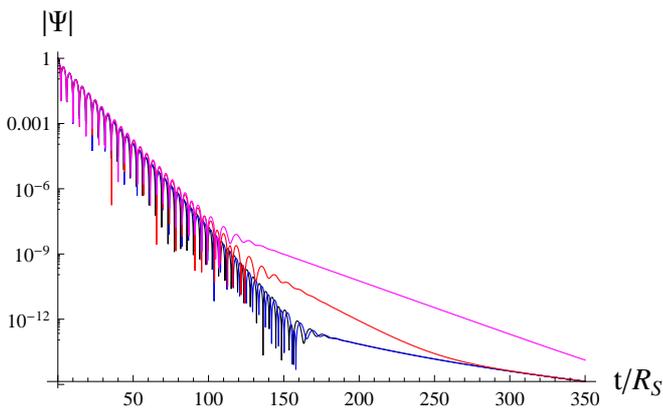}
\caption{Ringdown of the axial gravitational perturbations ($l=2$) for the Schwarzschild black hole (black) and for the Schwarzschild stars: $R=1.04R_{S}$ (blue, lower), $R=1.08R_{S}$ (red), $R=1.1R_{S}$ (magenta, upper).}\label{fig:profiles}
\end{figure}

Therefore, $r_{*}\to\infty$ corresponds to spacial infinity and $r_{*}\to-\infty$ when approaching the singular surface $r\to R_{0}^{+}$.

Note that in the vacuum exterior ($r\geq R$) $\epsilon=0$ and $p(r)=0$, hence the perturbation potential \eqref{potential} reduces to the Regge-Wheeler~\cite{regge1957} form
\begin{equation}\label{regge}
V^{out}_{l}(r) = \left(1-\frac{R_{S}}{r}\right)\left(\frac{l(l+1)}{r^2}-\frac{3R_{S}}{r^3}\right),\quad r\geq R.
\end{equation}
Here $R_{S}=2M=2m(R)$, where $M$ corresponds to the total mass of the star.

\begin{table}
\begin{tabular}{|c|c|c|}
\hline
$R/R_{S}$ & $\omega_i R_{S}$ & $\omega_l R_{S}$ \\
\hline
 $1.00$ & $0.7474-0.1779 \imo$& \\
 $1.01$ & $0.7486-0.1818 \imo$& \\
 $1.02$ & $0.7430-0.1831 \imo$& \\
 $1.03$ & $0.7398-0.1811 \imo$& \\
 $1.04$ & $0.7379-0.1777 \imo$& \\
 $1.05$ & $0.737~-0.175 \imo~$& \\
 $1.06$ & $0.737~-0.173 \imo~$&$-0.11\imo~$ \\
 $1.07$ & $0.737~-0.170 \imo~$&$-0.10\imo~$ \\
 $1.08$ & $0.737~-0.168 \imo~$&$-0.08\imo~$ \\
 $1.09$ & $0.737~-0.167 \imo~$&$-0.070\imo$ \\
 $1.10$ & $0.737~-0.166 \imo~$&$-0.057\imo$ \\
 $1.11$ & $0.737~-0.164 \imo~$&$-0.042\imo$ \\
 $1.12$ & $0.737~-0.163 \imo~$&$-0.023\imo$ \\
\hline
 $1.13$ &  \multicolumn {2}{c|}{$0.218055-2.4\times10^{-9} \imo$}\\
 $1.14$ &  \multicolumn {2}{c|}{$0.371288-1.2\times10^{-6} \imo$}\\
 $1.15$ &  \multicolumn {2}{c|}{$0.470264-2.6\times10^{-5} \imo$}\\
\hline
\end{tabular}
\caption{Dominant quasinormal frequencies of axial gravitational perturbations ($l=2$) obtained by fitting time-domain profiles. Initially the ringdown is governed by the oscillating mode $\omega_i$, at late times purely imaginary mode $\omega_l$ dominates. The last three lines correspond to dominant frequencies of the constant density star above the Buchdahl limit, taken from \cite{chandra1991b}.}\label{tabl:qnms}
\end{table}

In figure~\ref{fig:potp} we show the behavior of the effective potential as a function of the tortoise coordinate: at the star's surface it has a discontinuity due to discontinuity of  $\epsilon$. Since $p(r)$ grows unboundedly as $r\to R_{0}$ the potential has a negative gap from the righthand side of the peak, approaching $0^{-}$ as $r_*\to-\infty$. As the radius of the star approaches the Schwarzschild radius, the discontinuity and the negative gap become smaller and can be neglected, if $R$ is sufficiently close to $R_{S}$ (see Fig.~\ref{fig:potentials}).

In order to produce the time-domain profiles, we integrate the wavelike equation \eqref{master} rewritten in terms of the light-cone variables $u = t - r_*$ and $v = t + r_*$. The discretization scheme was suggested in~\cite{Gundlach:1993tp} and used in a number of papers (see for instance~\cite{Konoplya:2019hml,Konoplya:2011qq} and references therein) showing very good accuracy. It has the following form:
\begin{eqnarray}\label{eq:scheme}
\Psi(N) &=& \Psi(W) + \Psi(E) -
\Psi(S)  \\
& &- \Delta^2\frac{V(W)\Psi(W) + V(E)\Psi(E)}{8} +
\mathcal{O}(\Delta^4)\ ,
\nonumber
\end{eqnarray}
where we have used the following definitions for the points: $N = (u + \Delta, v + \Delta)$, $W = (u + \Delta, v)$, $E = (u, v + \Delta)$ and $S = (u,v)$. The initial data are specified on the two null surfaces $u = u_{0}$ and $v = v_{0}$.

From figure~\ref{fig:profiles} we see that the ringdown of Schwarzschild stars is similar to the quasinormal ringing of the Schwarzschild black hole and becomes indistinguishable as the star's radius approaches the Schwarzschild radius. For larger Schwarzschild stars ($R\lesssim 9R_{S}/8$), the purely imaginary mode appears at late time.
A similar phenomenon was observed in the Einstein-Gauss-Bonnet theory, where a purely imaginary nonperturbative mode appears in the spectrum~\cite{Konoplya:2008ix}, leading to instability for sufficiently large coupling~\cite{Gonzalez:2017gwa,Konoplya:2017zwo}. Although the effective potential for the gravitational perturbations of Schwarzschild stars also has a negative gap, it does not become deeper as $l$ grows, so that we do not have any indications of the eikonal instability~\cite{Cuyubamba:2016cug}. In addition, we have observed that for $l>2$ the imaginary mode does not dominate at late time.

Note, that the quasinormal spectrum of the Schwarzschild black hole has also the purely imaginary mode called \emph{algebraically special}. However, the algebraically special mode of the Schwarzschild black hole corresponds to high imaginary part, so that it does not dominate in the signal and, consequently, we cannot see it in the time-domain profile. The algebraically special mode for the Schwarzschild solution has the form~\cite{Chandrasekhar:1976zz}
\begin{equation}\label{algebraic}
\omega_{a}R_{S} = -\imo \frac{(l-1)l(l+1)(l+2)}{6R_{S}},
\end{equation}
so that the algebraically special mode for $l=2$ is $\omega = -4\imo/R_{S}$ what does not exclude possibility that the purely imaginary mode may go over into the known algebraically special mode in the Schwarzschild limit. If so, we should assume that the purely imaginary mode, which we observe here for $l=2$, exists also for higher $l$, but it does not dominate at late times for $l>2$ even when $R\to 9R_{S}/8$, because at higher $l$ the algebraically special mode \eqref{algebraic} is strongly damped.

After the ringdown phase and the transition consisting of exponential decay we see the asymptotic power-law tails, which are identical to those for the Schwarzschild black hole \cite{Price:1971fb,Price:1972pw},
\begin{equation}
| \Psi | \sim t^{- (2l +3)}.
\end{equation}

We also notice that the ringdown of the Schwarzschild star above the Buchdahl limit is qualitatively different (see Table~\ref{tabl:qnms}). The dominant mode has a small imaginary part and no purely damped modes were found in the spectrum~\cite{chandra1991b,Kokkotas1994}.

Summarizing this section, we can conclude that the Schwarzschild star can be a very good mimicker of Schwarzschild black hole, when measuring $l>1$ multipoles, as not only the ringdown phase is unaffected by the star interior, but also the expected echoes are absent, while the asymptotic tails are identical to those for the Schwarzschild ones.

\section{Conclusions}\label{sec4}

We have shown that the Schwarzschild stars are stable not only against spherical perturbations, but also against axial nonspherical ones, which is confirmed by the decayed time-domain profiles for the corresponding perturbation equations. This is one more step to establishing the viability of the Schwarzschild star model. At the same time, the Schwarzschild star can be a very good ‘mimicker' of the gravitational wave response of a black hole at $l=2$ and higher multipoles, because it can approach the Schwarzschild spectrum as closely as necessary. The latter is possible because the internal surface with an infinite redshift provides the same boundary conditions for quasinormal mode problem as a black hole does. Moreover, at later times, when gravitational echoes should show whether we have a gravastar or a classical black hole, the purely imaginary damping mode is dominating, instead of echoes, which goes over into the usual Schwarzschild asymptotic power-law tails.

The present study raises  a number of interesting questions. First of all, whether the polar gravitational perturbations are stable as well. In case of complete stability, the rotating model must be constructed beyond the linear order in rotation parameter and its stability, quasinormal modes and shadows  investigated. Another question is whether the purely imaginary modes which we found belong to the class of algebraically special modes. The latter can be checked by numerical calculations of quasinormal modes in the frequency domain, for example, with the help of the Frobenius method and accurate extrapolation to the Schwarzschild limit. Finally, an important question is whether the lower multipoles ($l=0,1$) could be seen in the gravitational wave signal we observe nowadays. If so, then this could be the way to distinguish the Schwarzschild star from a black hole.

\begin{acknowledgments}
This paper was supported by the grant 19-03950S of Czech Science Foundation (GAČR) and the “RUDN University Program 5-100”.
\end{acknowledgments}

\bibliography{main}
\bibliographystyle{iopart-num}
\end{document}